\documentclass[useAMS,usenatbib,fleqn]{mn2e}

\usepackage{verbatim}
\usepackage{graphicx}
\usepackage{amssymb}
\usepackage{amsmath}
\usepackage{xspace}
\usepackage[figuresright]{rotating}
\usepackage{pict2e}

\newcommand{\ro}[1]{\ensuremath{\textrm{#1}}}

\newcommand{\kmsec}{\ensuremath{~\ro{km}~ \ro{s}^{-1}}}
\newcommand{\ergs}{\ensuremath{~\ro{erg s}^{-1}}}
\newcommand{\cm}{\ensuremath{~\ro{cm}^{-2}}}
\newcommand{\Msol}{\ensuremath{M_{\odot}}}
\newcommand{\lya}{\ensuremath{\ro{Ly}\alpha}\xspace}
\newcommand{\ps}{Press-Schechter }
\newcommand{\st}{Sheth-Tormen }
\newcommand{\hi}{\ensuremath{\ro{H\textsc{i}}} }

\newcommand{\df}{\ensuremath{~ \ro{d} }}
\newcommand{\dd}{\ensuremath{\ro{d} }}

\newcommand{\cN}{\ensuremath{\mathcal{N}} }

\title[Emission and absorption properties of DLAs]{A joint model for the
emission and absorption properties of damped \lya absorption systems}

\author[Barnes \& Haehnelt]{Luke A. Barnes\thanks{E-mail:
lab@ast.cam.ac.uk (LAB); haehnelt@ast.cam.ac.uk (MGH)} and Martin G. Haehnelt\\ 
 Institute of Astronomy, Madingley Road, Cambridge, CB3
0HA}

\begin{document}

\date{not yet submitted}

\pagerange{\pageref{firstpage}--\pageref{lastpage}} \pubyear{2008}

\maketitle 

\label{firstpage}

\begin{abstract} 
The recently discovered population of ultra-faint extended line 
emitters, with fluxes of a few times $ 10^{-18} \ergs \cm$ at $z\sim 3$, can 
account for the majority of the incidence rate of Damped \lya systems (DLAs) 
at this redshift if the line emission is interpreted as \lya. We show here that 
a model similar to that proposed by \citet{2000ApJ...534..594H}, which explains
the incidence rate and kinematics of DLAs in the context of $\Lambda$CDM models 
for structure formation, also reproduces the size distribution of the new 
population of faint \lya emitters for plausible parameters. This lends further 
support to the interpretation of the emission as \lya, as well as the identification 
of the emitters with the hitherto elusive population of DLA host galaxies. 
The observed incidence rate of DLAs together with the observed space density 
and size distribution of the emitters suggest a duty cycle of $\sim 0.2-0.4$ 
for the \lya emission from DLA host galaxies. We further show that \lya cooling 
is expected to contribute little to the \lya emission for the majority of emitters. This 
leaves centrally concentrated star formation at a rate of a few tenths 
\Msol yr$^{-1}$, surrounded by extended \lya halos with radii up to 30-50 kpc, 
as the most plausible explanation for the origin of the emission. Both the 
luminosity function of \lya emission and the velocity width distribution of 
low ionization absorption require that galaxies inside Dark Matter (DM) halos 
with virial velocities $\la 50-70 \kmsec$ contribute little to the incidence rate
of DLAs at $z\sim 3$, suggesting that energy and momentum input due to star formation 
efficiently removes gas from these halos. Galaxies with DM halos with virial 
velocities of $100-150 \kmsec$ appear to account for the majority of DLA host 
galaxies. DLA host galaxies at $z\sim 3$ should thus become the building blocks 
of typical present-day galaxies like our Milky Way.
\end{abstract}


\section{Introduction} 
Quasar absorption spectra provide excellent
probes of the distribution of baryons in the high-redshift
Universe. Damped Lyman Alpha systems (DLAs, historically defined as
having a neutral hydrogen column density $N_{\ro{HI}} > 2 \times
10^{20}$cm$^{-2}$) are particularly useful as they are likely to play
an important role as a reservoir of gas for the formation of stars and
galaxies at high redshift. They dominate
the neutral gas content of the Universe between $z \sim 0-5$, and at $
z \sim 3.0-4.5$ their neutral gas content is comparable to visible
stellar mass in present-day galaxies
\citep{1986RSPTA.320..503W,1996MNRAS.283L..79S,2000ApJ...543..552S}. 
DLAs thus form an important link between primordial plasma and the 
stellar structures that form from it.

In spite of observations of over 1000 DLAs,
\citet{2005ARA&A..43..861W} still conclude that the question ``what is
a damped \lya system?'' has not yet been answered conclusively. One of the reasons
for this is that absorption spectra provide only indirect 
information in velocity space, and only probe the gas along one line-of-sight through the galaxy. 
The \lya absorption feature itself provides no
information about the velocity structure of the DLA, because of the
large optical depth even in the damping wings. This has led observers
to look at the associated low ionization metal absorption
features. Low ionization species like SiII, CII and FeI are believed
to be good tracers of the neutral gas in
DLAs. \citet{1997ApJ...487...73P} developed the velocity width
distribution of metal absorption features into an important diagnostic tool for
DLAs. The much lower absorption optical depth allows us to extract
detailed kinematical information for the gas in DLAs. The
absorption profiles are normally clumpy and asymmetric, with the
strongest absorption feature often occurring at one edge of the
profile. Velocity widths range from 30$\kmsec$ up to several hundred
$\kmsec$. Note, however, that there is a distinctive lack of very
narrow absorption profiles.

Kinematical models aiming to reproduce the velocity width data fall
into two categories. \citet{1986RSPTA.320..503W} suggested a
close connection between DLAs and disks of present-day spiral
galaxies. \citet{1997ApJ...487...73P} modelled DLAs as thick rotating disks
with a rotation speed ($\sim 200\kmsec$) typical of
present-day galaxies. \citet{1998ApJ...495..647H} challenged this
interpretation and demonstrated that it is not unique. The merging of proto-galactic
clumps expected in Cold Dark Matter (CDM)-like models for structure formation can explain
the shape of the profiles equally well. Galactic winds are also likely
to play an important role \citep[e.g.][]{1999MNRAS.308L..39F}.
They further showed that for
the same virial velocity width, merging clumps produce significantly
larger velocity width and argued that the latter interpretation is
favoured by the observed velocity width distribution in the context 
of the CDM paradigm for structure formation \citep{2000ApJ...534..594H}. 

The level of enrichment with metals provides another important clue as 
to the nature of DLAs --- see \citet{2004cmpe.conf..257P,
2006fdg..conf..319P,2005ARA&A..43..861W} for reviews. Significant
metal absorption is found in all DLAs, though
as a population they are metal poor. DLAs are also relatively dust
free \citep{2001A&A...379..393E,2006MNRAS.367..211W}. Initially, the low metallicity was 
used to argue that DLA host galaxies are the chemically
unevolved but otherwise very similar counter-parts of typical 
present-day spiral galaxies \citep{1986RSPTA.320..503W}. In the model of 
\citet{1998ApJ...495..647H}, DLAs instead preferentially probe the 
outer parts of much less massive galaxies, many of which end up 
as building blocks of typical present-day galaxies that form by
hierarchical merging in CDM-like models for structure formation. 
Recently, \citet{2008arXiv0804.4474P} demonstrated that such 
a model fits the observed metallicity of DLAs at $z\sim 3$ 
very well.

Models of DLAs invariably predict more than quasar absorption features
can in principle test. As already mentioned, absorption is measured along a single line of
sight and thus is a one-dimensional probe of the properties of the
DLA. To explore the spatial extent and structure of the DLA, we need
to observe DLAs in emission. Attempts to do this have focussed on both
line and continuum emission. \lya emission holds great
potential in this respect. Star-formation, 
cooling radiation and fluorescent re-emission of the meta-galactic 
UV background are all expected to contribute at different levels. In
addition, stellar continuum emission from the newly formed stars
should also be bright enough to be detectable.

Many observers have attempted to find the galaxy counterparts of DLAs
at high redshift in emission by searching adjacent to quasar
sightlines with known absorption
systems \citep{1999MNRAS.305..849F,1999MNRAS.309..875B,2000ApJ...536...36K,
2001ApJ...551...37K,2001MNRAS.326..759W,2007A&A...468..587C}. This is
a difficult task, as the light of the extremely bright quasar must
be accurately subtracted to study the light of the galaxy, which is very
faint in comparison. \citet{2000ApJ...536...36K} and others caution of the
possibility that a given emission feature is a Point Source Function
(PSF) artefact rather than a real source. 
\citet{2007A&A...468..587C} report that, for $z > 2$, six DLA
galaxies have been confirmed through spectroscopic observation of \lya emission,
with other techniques producing a few additional
candidates. \citeauthor{2007A&A...468..587C} added another six \lya
emission candidates to this group. A quantitative statistical
interpretation of the many (largely unsuccessful) searches is difficult
if not impossible, but the rather low success rate appears to be
consistent with their interpretation as galaxies of rather low mass
and star formation rate. 
 
\citet{2008ApJ...681..856R} reported the results of a long-slit search
for low surface brightness \lya emitters at redshift $z \sim 3$, which 
reached flux levels that are about a 10 times lower than previous \lya surveys
at this redshift. They found 27 faint line emitters, many of which are extended in wavelength
and real space. They argue that the majority of the emitters are
likely to be \lya (rather than low-$z$ interlopers), powered
by a central region of star formation and processed by radiative
transfer through surrounding neutral
hydrogen. \citet{2008ApJ...681..856R} note that the incidence
rate inferred from the space density and the size distribution of the
emitters is similar to that of DLAs, and suggest that they are the host population of DLAs and high column density Lyman
Limit Systems (LLS). If this is indeed the case then the observations of 
\citet{2008ApJ...681..856R} give us --- for the first time --- the 
size distribution and space density of DLA host galaxies 
\citep[Figure 19 of ][]{2008ApJ...681..856R}.

Models predicting the velocity width and size distribution of DLAs have been
constructed based on the observed luminosity function of galaxies \citep{1999MNRAS.305..849F,2008ApJ...683..321F} and on the Press-Schechter formalism
\citep{1974ApJ...187..425P} in conjunction with numerical
simulations
\citep{1997ApJ...484...31G,1997ApJ...486...42G,1998ApJ...495..647H,
2000ApJ...534..594H,2004MNRAS.348..421N,2007ApJ...660..945N}. More
recently numerical simulations have attempted to model the entire DLA
population self-consistently \citep{2006ApJ...645...55R,2007arXiv0710.4137R,2008arXiv0804.4474P}.

In this paper, we revisit models for the absorption properties of DLAs
in the light of the size distribution data of
\citet{2008ApJ...681..856R}, improved data on the velocity width
distribution from metal absorption lines (which has presented a
challenge to purely numerical simulations --- see
\citet{2007arXiv0710.4137R}) and numerical simulations
with increased resolution, box size and
sophistication incorporating the
additional physics of radiative transfer, gas chemistry and star
formation.

Throughout this work we use cosmological parameters of 
the 5 year WMAP data \citep{2008arXiv0803.0732H}: $(h, \Omega_M, \Omega_b,
\Omega_{\Lambda}, \sigma_8, n) = (0.701, 0.279, 0.046, 0.721, 0.817,
0.96)$.


\section{A joint model for the kinematical properties of DLAs and the
cumulative size distribution of the faint \lya
emitters} \label{dlamodel}

\subsection{The Haehnelt et al. model}

\citet{2008ApJ...681..856R} argued that their population of faint \lya emitters is
the same or has at least a large overlap with that of DLA/LLs host
galaxies. They further pointed out that the space density and sizes
should agree well with those predicted by the DLA model of
\citet{1998ApJ...495..647H}, which models DLAs in the context of CDM
models of structure formation.

We here revisit and update this model to investigate
whether it can explain the properties of DLAs and the
new population of faint emitters, assuming that these are the same
objects. We start with summarizing the salient properties of the
model. As discussed in the introduction, \emph{ab initio} numerical
simulation of the gas at the centre of galaxies, where complex
non-linear gas physics including star formation and the associated
feedback are important, is still very challenging. The
Haehnelt et al. model therefore takes a hybrid approach, using a
combination of Press-Schechter formalism and results from numerical
simulations to model the kinematic properties of DLAs 
(see \citet{2006MNRAS.371.1519J} for a 
semi-analytical model of DLAs that explicitly models feedback).

The model uses the space density of DM halos as a proxy for the space
density of DLA host galaxies. With the refinement to the \ps formalism
introduced by \citet{2002MNRAS.329...61S}, the number of dark matter
halos per unit comoving volume at redshift $z$ with mass (baryonic +
CDM) in the interval $(M,M+\dd M )$ can be estimated as,
\begin{equation}\label{eq:nst} n_M(M,z) \df M = A \left( 1 +
\frac{1}{\nu'^{2q}} \right) \sqrt{\frac{2}{\pi}} \frac{\rho_{0}}{M}
\frac{\dd\nu'}{\dd M} \exp \left(-\frac{\nu'^2}{2} \right) \df M,
\end{equation} where $\sigma_M$ is the rms fluctuation amplitude of the
cosmic density field in spheres containing mass $M$, $\rho_{0}$ is the
present cosmic matter (baryonic + CDM) density, $\nu' = \sqrt{a}\nu$,
$\nu = \delta_c / [D(z) \sigma_M]$. $D(z)$ is the growth factor at
redshift $z$ \citep{1992ARA&A..30..499C}, $\delta_c = 1.686$, $a =
0.707$, $A \approx 0.322$ and $q = 0.3$. We have used the fitting
formula in \citet{1999ApJ...511....5E} to calculate the matter power
spectrum.

\begin{figure*}
	\centering
	\begin{minipage}[c]{0.47\textwidth}
		\centering 
 		\includegraphics[width=\textwidth]{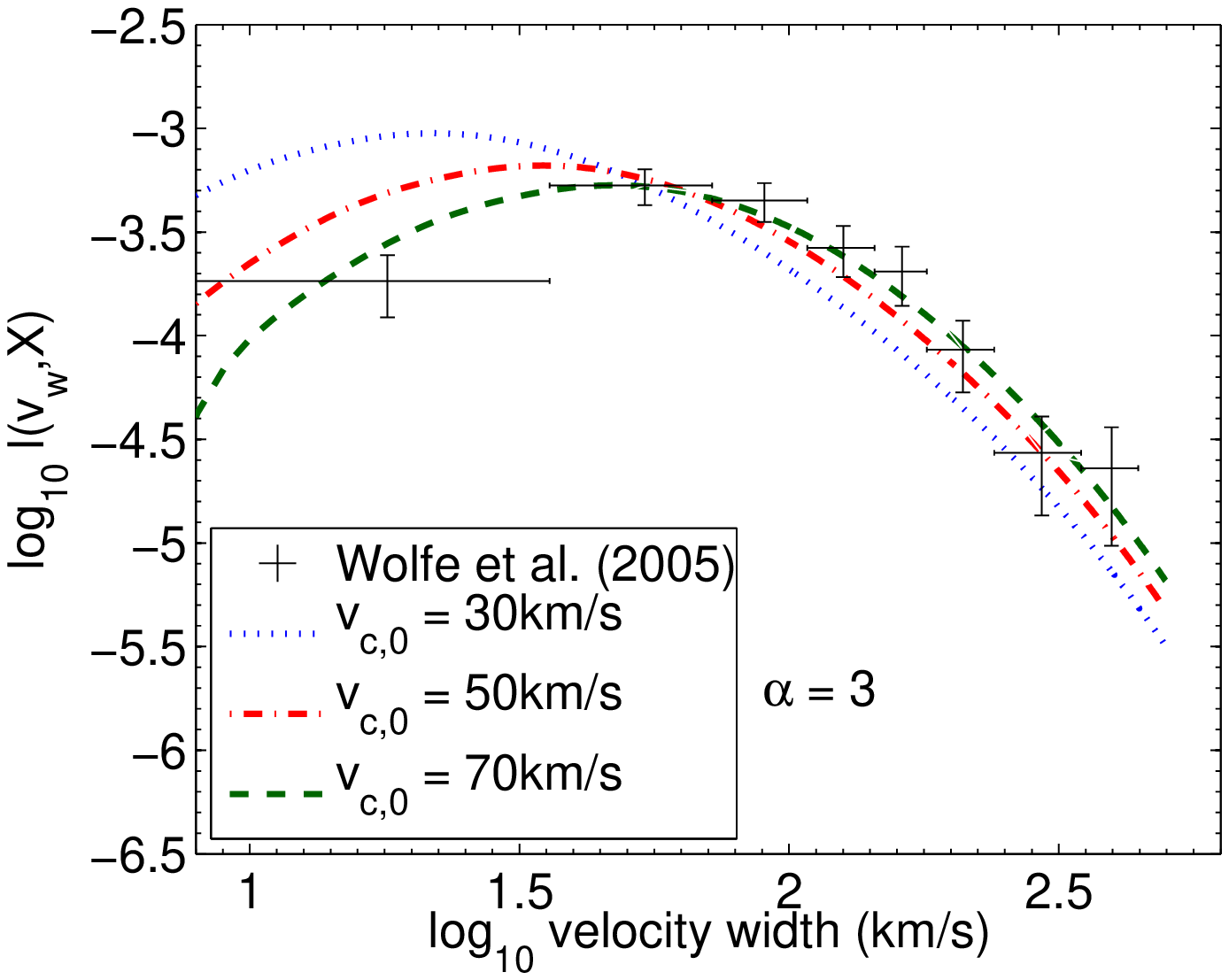}
	\end{minipage}
	\begin{minipage}[c]{0.47\textwidth}
		\centering 
 		\includegraphics[width=\textwidth]{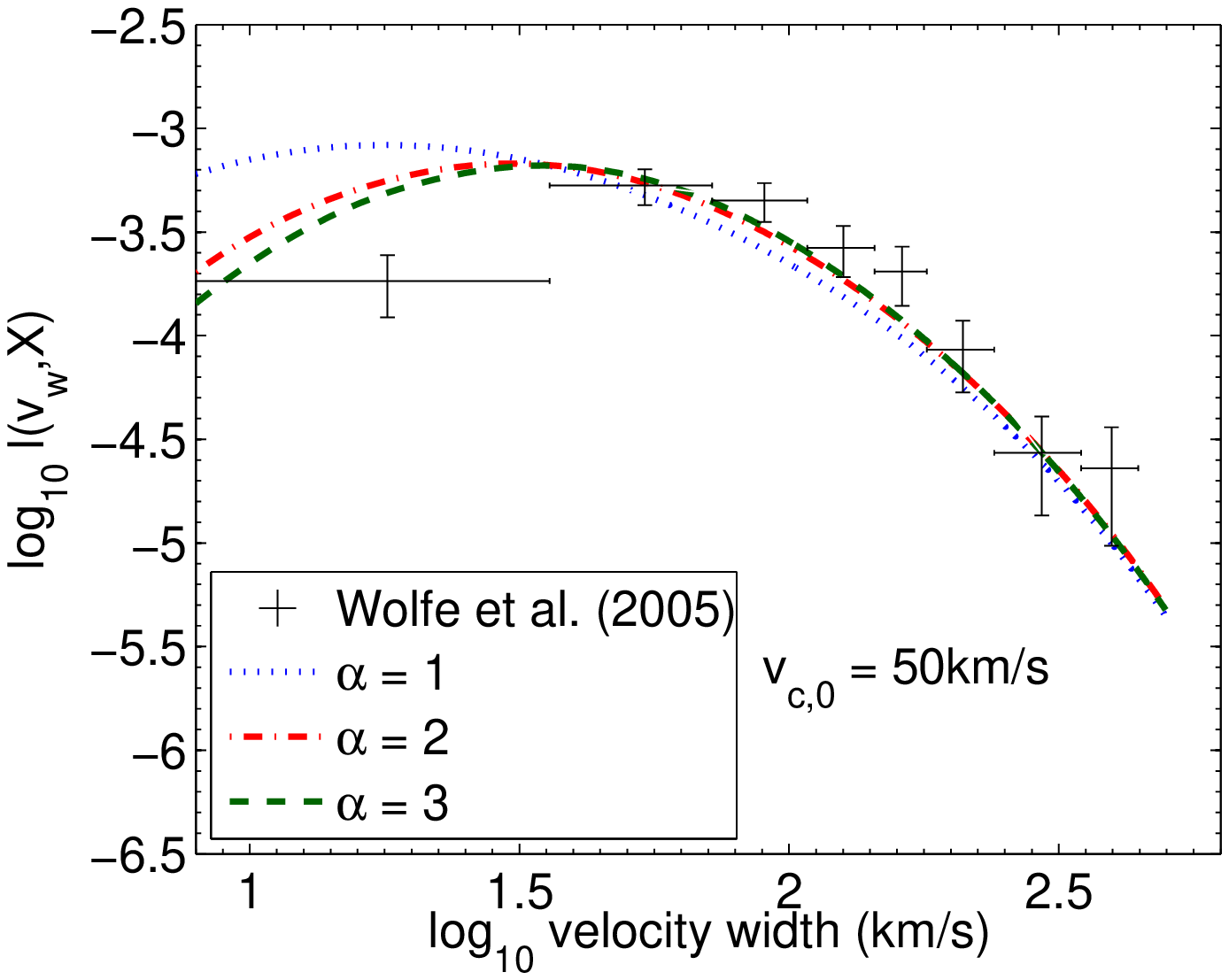}
	\end{minipage}
	\begin{minipage}[c]{0.47\textwidth}
		\centering 
 		\includegraphics[width=\textwidth]{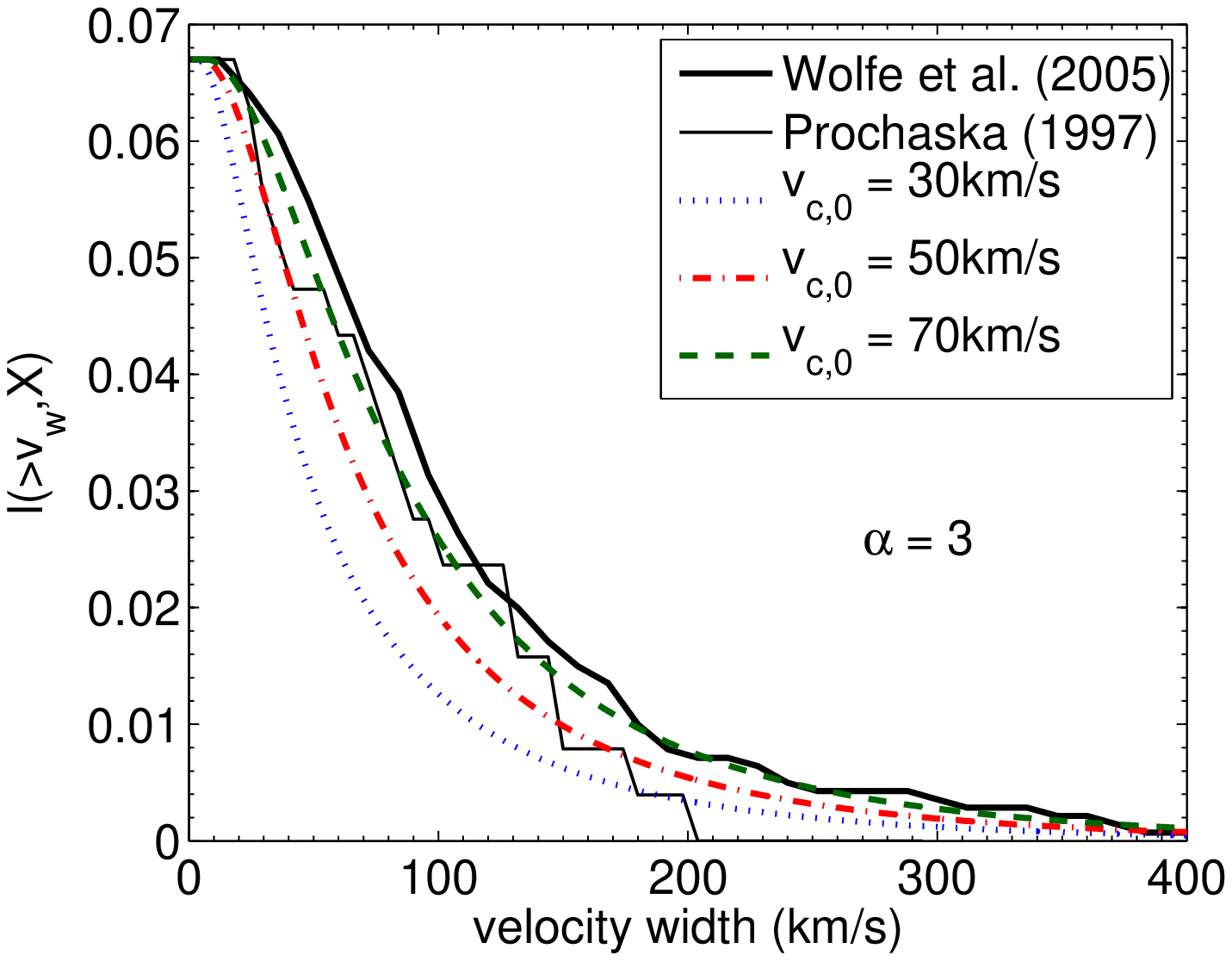}
	\end{minipage}
	\begin{minipage}[c]{0.47\textwidth}
		\centering 
 		\includegraphics[width=\textwidth]{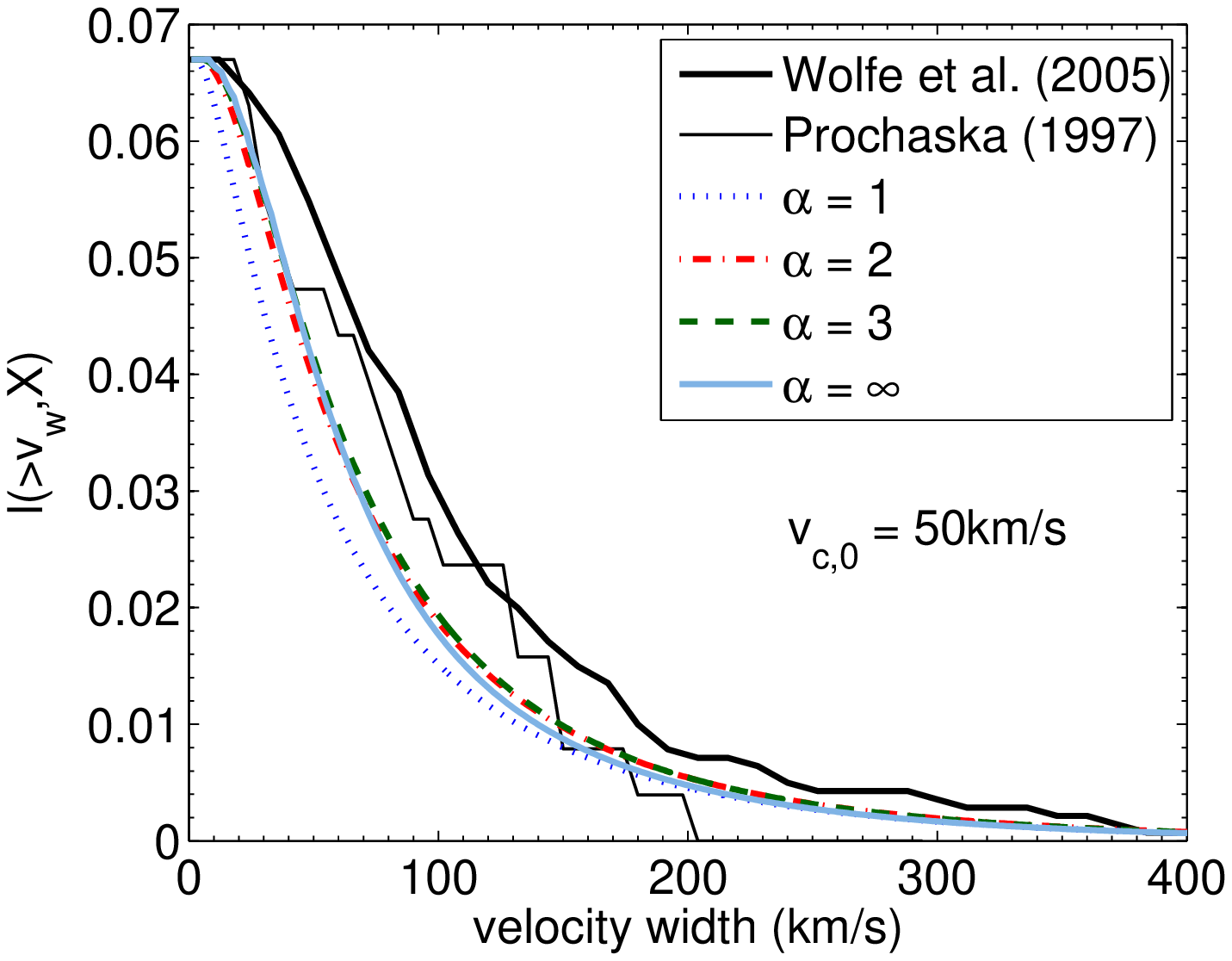}
	\end{minipage}
 \caption{\emph{Top Panels:} The velocity width distribution $l(v_\ro{w},X)$ of the
associated low-ionization metal absorption of DLAs. The black crosses
show the observational data compiled in \citet{2005ARA&A..43..861W}.
\emph{Bottom Panels:} The cumulative velocity width distribution $l(>v_\ro{w},X)$ of the
associated low-ionization metal absorption of DLAs. The thick solid
black curve shows the observational data of
\citet{2005ARA&A..43..861W}. The thin, solid black curve shows the
older observational data of \citet{1997ApJ...487...73P}.
\emph{Left Panels:}
The velocity width distribution of our model for an exponential
suppression at small virial velocities of the cross section for damped \lya absorption as given
by Equation \eqref{eq:sigvc2} with $\alpha = 3$ and $v_\ro{c,0} = 30, 50,
70 \kmsec$, respectively. The cross section normalization for these
models as given by Equation \eqref{eq:dndz} is characterised by $r_0 =
18.5, ~23.5, ~28.4$ kpc.
\emph{Right Panels: } The velocity width
distribution for $v_\ro{c,0} = 50 \kmsec$ and a range of values
of $\alpha = 1,2,3, \infty$. For these models, $r_0 = 21.9, ~23.6, ~23.9,
~23.5$ kpc.}
 \label{fig:plotw}
\end{figure*}

We will be interested in the kinematic properties of DLAs, for which the
virial velocity is a more convenient quantity to characterize the DM
halo than the mass. The two are related as follows
\citep[e.g.][]{2004MNRAS.355..694M},
\begin{equation} \label{eq:vc} \begin{split} v_{\ro{c}} = 106 \kmsec
&\left( \frac{\Delta_{\ro{v}}}{174} \right)^{\frac{1}{6}} \left(
\frac{\Omega_M h^2}{0.137} \right)^{\frac{1}{6}} \\ & \times \left(
\frac{1+z}{4} \right)^{\frac{1}{2}} \left( \frac{M}{10^{11} \Msol}
\right)^{\frac{1}{3}},
\end{split} \end{equation} where $\Delta_{\ro{v}}$ is the overdensity
of the halo \citep[see][]{1998ApJ...495...80B}.

We use a simple power-law relation between virial velocity of DM
halo and absorption cross section
\begin{equation} \label{eq:sigvc1} \sigma(v_{\ro{c}}) = \pi r_0^2
\left(\frac{v_{\ro{c}}} {200 \kmsec} \right)^\beta
\end{equation}
where $\beta$ and $r_0$ are parameters. The value of
$\beta$ has been the source of some
controversy, due to numerical simulations still finding it challenging to 
reliably model the spatial distribution of the gas in the high-density region
probed by DLAs and LLs. \citet{1997ApJ...484...31G} originally favoured a value
 of $\beta = 2.94$ (at $z = 3$, for a $\Lambda$CDM cosmology), but
their later work revised this to $\beta =
1.569$. \citet{2001ApJ...560L..33P} pointed out that this low value of
$\beta$ is incompatible with the observed DLA velocity widths. \citet{2000ApJ...534..594H}
found that the observed velocity width distribution of metal-lines
is reproduced well with a value of $\beta \sim 2.5$, which we also use here.

Haehnelt et al. further found that they needed to introduce a lower
cut-off in virial velocity in order to fit the velocity width
distribution of low-ionization absorption systems. Otherwise their
model predicted too many very narrow metal absorption systems, which
are not observed. They therefore assumed that
DM halos with virial velocities smaller than a minimum velocity
$v_{\ro{min}}$ do not host DLAs. 

Here we will slightly relax this assumption and model the
suppression of the cross-section of DLAs in halos with small circular
velocities as a gentler exponential decline, 
\begin{equation} \label{eq:sigvc2} \sigma(v_{\ro{c}}) = \pi r_0^2
\left(\frac{v_{\ro{c}}} {200 \kmsec} \right)^\beta ~ \exp \left( -
\left(\frac{v_\ro{c,0}}{v_{\ro{c}}} \right)^{\alpha} \right).
\end{equation} We consider a range of values for the parameters
$v_\ro{c,0}$ and $\alpha$. Note that a sharp cut-off corresponds to
$\alpha =\infty$.

It remains to choose the parameter $r_0$. We will follow
\citet{2000ApJ...534..594H} by fixing $r_0$ so that the overall rate
of incidence of absorbers per unit redshift $\dd \cN / \dd z$ agrees
with the observational value,
\begin{equation} \label{eq:dndz} \frac{\dd \mathcal{N}} {\dd z} =
\frac{\dd l_p} {\dd z} ~ (1+z)^3 ~ \int_0^\infty \sigma(M,z) ~
n_M(M,z) ~ \df M ~.
\end{equation} We take the value of $\dd \cN / \dd z = 0.24$ at $z =
3$, which is consistent with $\dd \cN / \dd X = 0.067$
\citep{2005ApJ...635..123P,2005MNRAS.363..479P}. The ratio of proper
distance interval to redshift interval is given by,
\begin{equation} \frac{\dd l_p}{\dd z} = \frac{c} {H(z) (1+z)} ~,
\end{equation}
and the so-called absorption distance, X, is defined by,
\begin{equation} \dd X \equiv \frac{H_0} {H(z)} (1+z)^2 \df z ~.
\end{equation} The resulting values of $r_0$ are given in the caption
of Figure \ref{fig:plotw}.

Further, it is useful to define the line density distribution ($N$),
such that the number of intersections ($\dd^2 \cN$) of a random line
of sight with systems that have mass in the interval $(M,M+\dd M )$,
located in the redshift interval $(z,z+\dd z )$ is,
\begin{align} \label{eq:NMz} \dd^2 \cN &\equiv N(M,z) \df M \df z \\
&= (1+z)^3 ~ n_M(M,z) ~ \sigma(M,z) ~ \frac{\dd l_p} {\dd z} ~\dd
M~\dd z.
\end{align}


\subsection{The velocity width distribution of low-ionization
absorption} To calculate the velocity width distribution,
Haehnelt et al. used a conditional probability distribution: $p(v_\ro{w}
| v_{\ro{c}}) \df v_\ro{w}$ is the probability that a DLA in a halo with circular
velocity $v_{\ro{c}}$ has a velocity width in the interval $(v_\ro{w},v_\ro{w}+\dd v_\ro{w}
)$. On the basis of their numerical simulations, they assumed it to be
a function of the ratio $v_\ro{w} / v_{\ro{c}}$. The number of systems 
along a random line of sight in the interval $(X,X+\dd X )$
with velocity width greater than $v_\ro{w}$ is given by
\begin{equation} l(>v_\ro{w},X) = \int_{v_\ro{w}}^{\infty} \left[
\int_{0}^{\infty} p(v_\ro{w} | v_{\ro{c}}(M))\frac{\dd^2 \cN}{\dd X ~ \dd M} \dd M
\right] \dd v_\ro{w}.
\end{equation} We will use here the conditional probability distribution
as given in \citet[][Figure 1]{2000ApJ...534..594H}. The distribution peaks at $v_\ro{w} / v_{\ro{c}} \approx 0.6$, dropping to zero below $v_\ro{w} / v_{\ro{c}} \approx 0.1$ and above $v_\ro{w} / v_{\ro{c}} \approx 2$. The numerical
simulations on which the distribution is based did not contain 
star formation feedback. Simulations have become
more sophisticated since then. We have therefore compared the
$p(v_\ro{w} | v_{\ro{c}})$ distribution used here with that from the simulations
of \citet[][and private communication]{2008arXiv0804.4474P}, which
incorporate the effects of star formation and supernovae on the kinematics and spatial
distribution of the gas in a simple manner. The differences in $p(v_\ro{w}
| v_{\ro{c}})$ are small. The resulting difference in $l(>v_\ro{w},X)$ is also small, at 
most $10 \%$. Unless there is a fortuitous cancellation of different effects, 
this suggests that star formation in the simulations has a small effect on 
$p(v_\ro{w} | v_{\ro{c}})$. This is somewhat surprising, given the significant
differences in resolution, cosmological volume and additional physics,
albeit reassuring for our modelling. Note, however, that the 
simulations still fail to produce realistic galactic winds, 
probably due to the rather simplistic fashion in which stellar
feedback is incorporated.

\begin{figure*} \centering 
\begin{minipage}[c]{0.47\textwidth} \centering
\includegraphics[width=\textwidth]{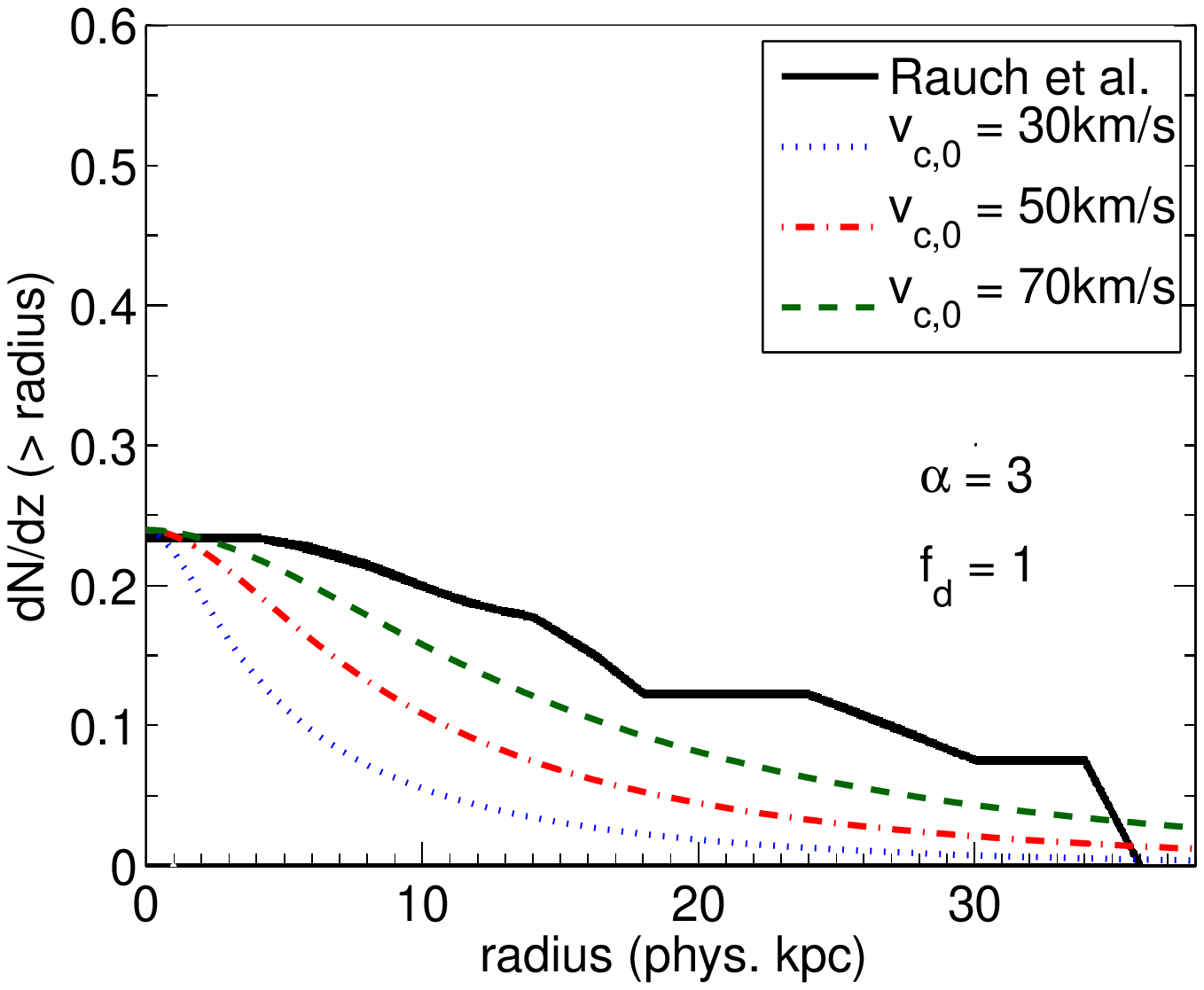}
\end{minipage} 
\begin{minipage}[c]{0.47\textwidth} \centering
\includegraphics[width=\textwidth]{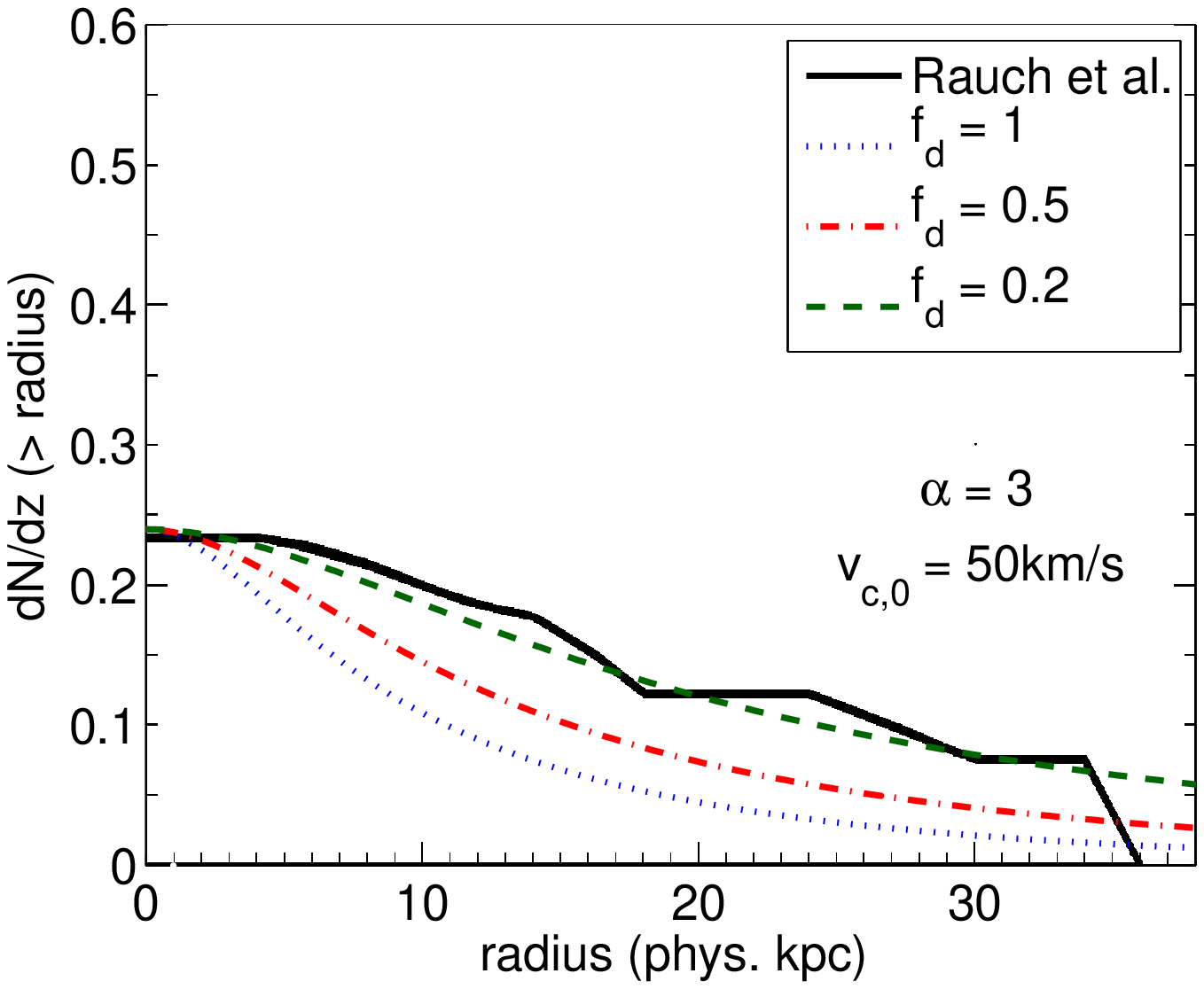}
\end{minipage}
\caption{The cumulative incidence rate inferred from the size
distribution of the population of faint \lya emitters. The solid black
curves are the observational data of
\citet{2008ApJ...681..856R}. \emph{Left:} Cumulative incidence rate in
our model for $\alpha = 3$ and a range of values of $v_\ro{c,0} = 30, 50,
70 \kmsec$. \emph{Right:} The cumulative incidence rate in our model
for $v_\ro{c,0} = 50 \kmsec$ and a range of values of the duty cycle $f_\ro{d} =
1,~0.5,~0.2$.
}
\label{fig:plot1}
\end{figure*}

Figure \ref{fig:plotw} compares the predicted velocity width
distribution with the observational data of
\citet{2005ARA&A..43..861W}. The data of \citet{1997ApJ...487...73P}
is also shown in the bottom panels and is reasonably well fit with $v_\ro{c,0}= 50
\kmsec$. This is very similar to what was found by
\citet{2000ApJ...534..594H}. The new compilation of velocity width
data by \citet{2005ARA&A..43..861W} extends to significantly larger
velocities and appears to require a somewhat larger value of $v_\ro{c,0}=
70 \kmsec$. The apparent lack of neutral gas in small dark matter
halos can be plausibly attributed to the feedback effects of star 
formation and/or photo-heating due to the meta-galactic UV background. 
In most numerical simulations and models of galaxy
formation, these feedback mainly affects halos
with somewhat smaller virial velocities than this. So either halos with small
virial velocities have larger velocity widths than we have
assumed here (i.e. $p(v_\ro{w} | v_{\ro{c}})$ is different), or else stellar feedback in halos with virial velocities of 
$v_\ro{c,0}= 50 \kmsec$ is more efficient than generally assumed.
The cumulative velocity width distribution
is not very sensitive to the shape of the cut-off for $\alpha \ge 2$.

The high velocity tail of the velocity distribution as compiled by
\citet{2005ARA&A..43..861W} has proven difficult to reproduce with
numerical simulations, which attempt to model the spatial distribution
of the neutral gas in DLAs self-consistently rather than assume a
scaling of the absorption cross section with the virial velocity of DM
halos. Generally, the simulations fail to produce a sufficient number
of absorption systems with velocity widths as wide as observed. This
is normally attributed by the authors of these studies to the fact
that momentum and energy input into the gas due to star
formation may not have been modelled with sufficient sophistication
\citep{2004MNRAS.348..421N,2007ApJ...660..945N,
2006ApJ...645...55R,2007arXiv0710.4137R,2008arXiv0804.4474P}.


\subsection{The cumulative size distribution}

We now ask whether our model, which successfully fits the velocity
width distribution of the associated low-ionization metal absorption
of DLAs, can also reproduce the cumulative size distribution of the
emission regions of the new population of faint \lya emitters, as shown by
black solid curve in Figure \ref{fig:plot1} \citep[][ Figure 19, \hi corrected]{2008ApJ...681..856R}. It
asymptotes to $\dd \cN / \dd z = 0.23$, which is very similar to the observed
incidence rate of DLAs at the same redshift.
To calculate the cumulative incidence
rate for the DLA host galaxies in our model, we must relate the size
of the emission region to the mass of the corresponding DM halo. We do
this simply by assuming that the absorption
cross-section is related to radius as if it was a sphere. The
cumulative incidence rate is then given by
\begin{equation} \frac{\dd \cN} {\dd z}(>r,z) = \int_{M(r)}^{\infty}
N(M',z) \df M'.
\end{equation} The result is shown\footnote{We do not show the impact of altering $\alpha$
here. The effect is small, especially beyond a radius of $10$kpc.} in the left panel of Figure
\ref{fig:plot1}. The
predicted sizes are consistently smaller than the observed sizes by a
factor of between 1.5 to 3, suggesting that the emission regions of
the \lya emitters are larger than the cross section for damped
absorption. Even though the integrated inferred incidence rate is
similar to that of DLAs, the emission regions of the \lya emitters can
thus not be identical to the regions responsible for damped \lya
absorption. Nevertheless, the two can be closely related.

\begin{figure*} \centering
\begin{minipage}[c]{0.48\textwidth} \centering
\includegraphics[width=\textwidth]{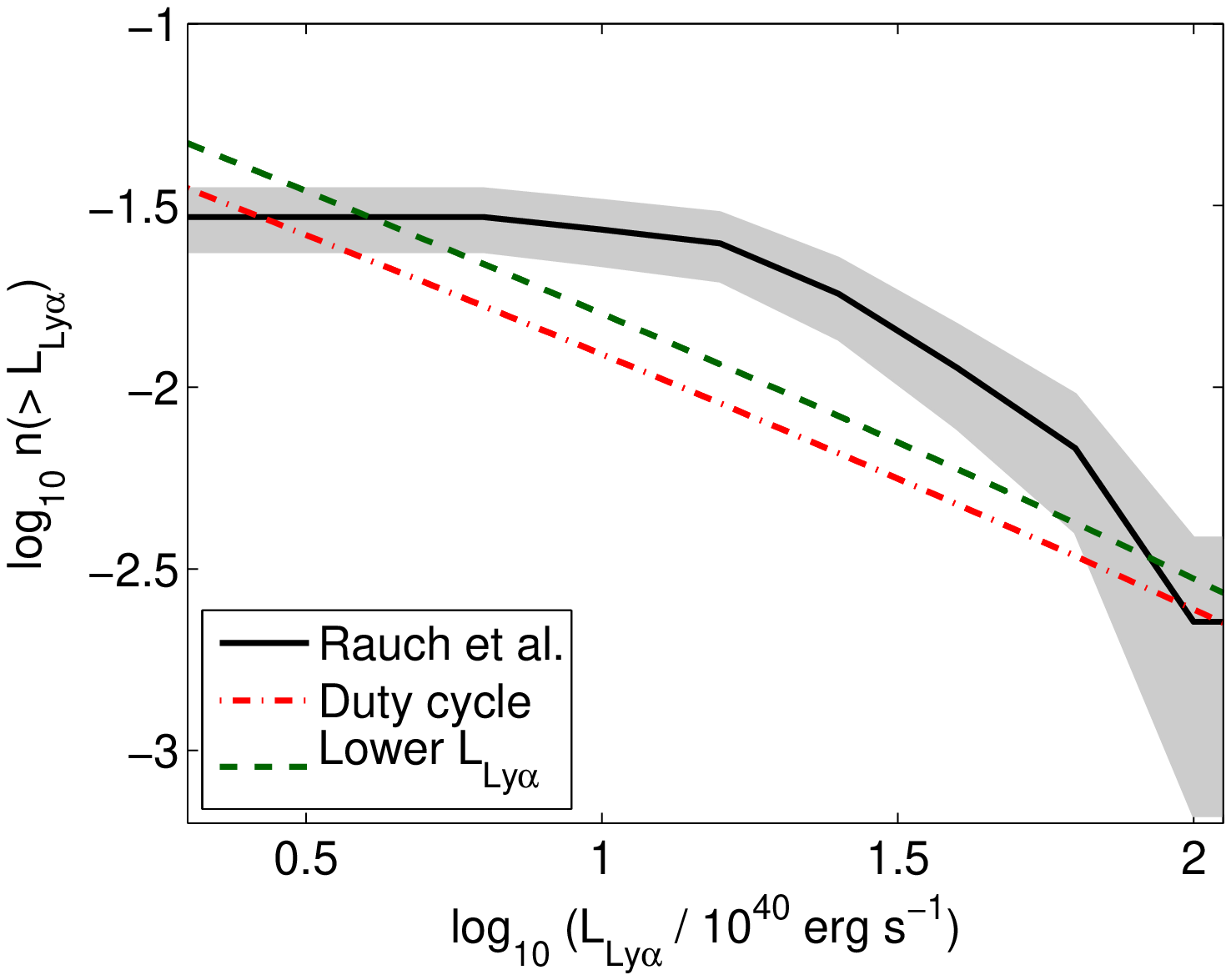}
\end{minipage} \hspace{0.03\textwidth}
\begin{minipage}[c]{0.48\textwidth} \centering
\includegraphics[width=\textwidth]{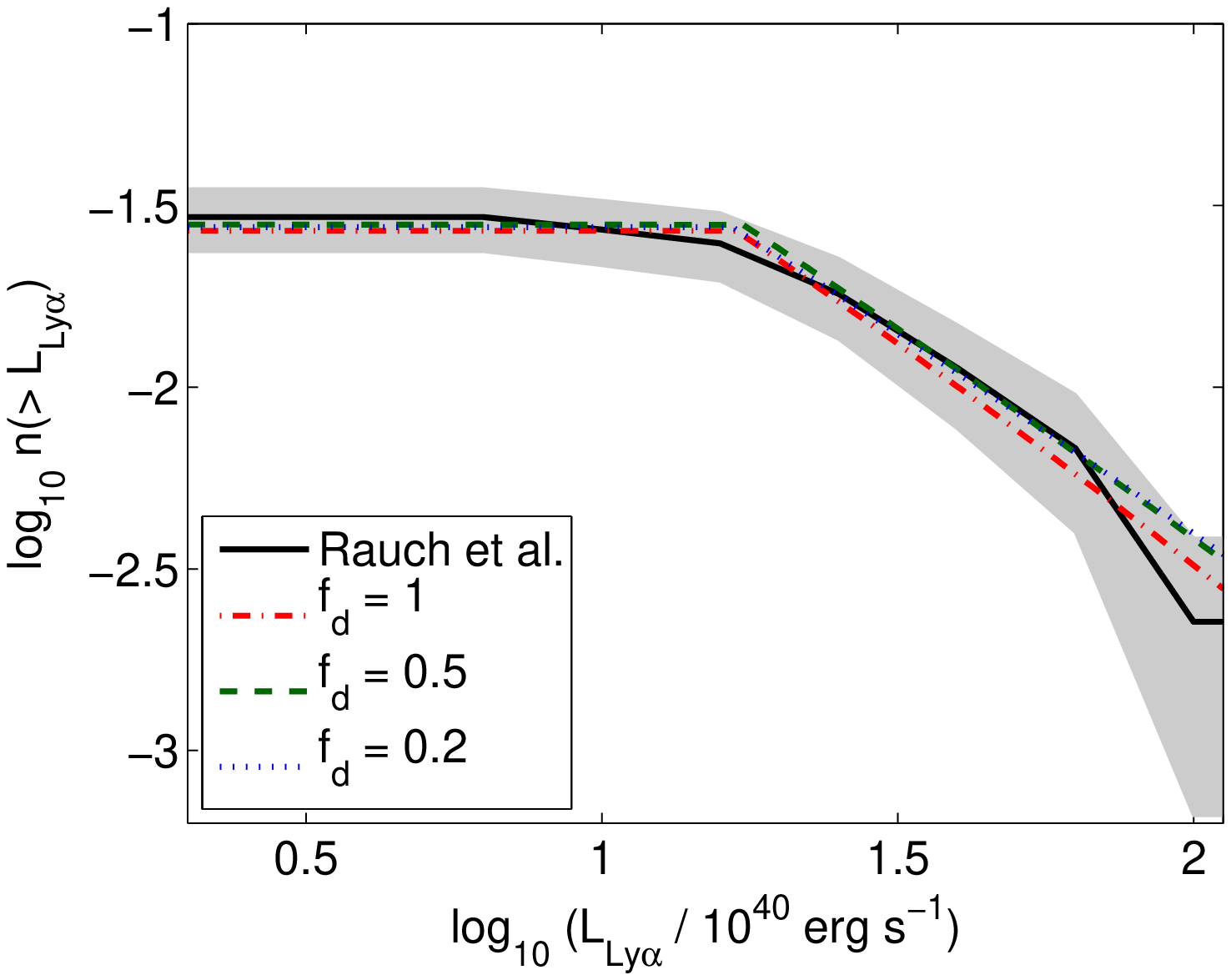}
\end{minipage}
\caption{The cumulative luminosity function of the \lya emitters. The
black solid curve in both panels is the data taken from Fig.~9 of
\citet{2008ApJ...681..856R}. The grey, shaded region is an estimate of the
1-$\sigma$ confidence interval. \emph{Left:} The blue solid curve and
the dashed red curves are for optimistic models for the \lya emission
due to \lya cooling radiation as described in the text. \emph{Right:}
A simple model for the \lya emission due to star formation as
described in the text. The relevant parameters for $f_\ro{d} =
(1,~0.5,~0.2)$ are $(L^{\ro{s}}_{0}, v_{\ro{min}} [\kmsec]) = (4
\times 10^{41}, 75), (8 \times 10^{41}, 60), (1.8 \times 10^{42},
45)$ respectively.}
\label{fig:cool}
\end{figure*}

To demonstrate this we now explore a simple model where only a
fraction $f_\ro{d}$ of the halos are emitting \lya radiation above the 
detection threshold at any one time. To keep the total inferred
incidence rate fixed, we allow the cross section of the individual
\lya emission regions to be larger than those for damped \lya
absorption by a corresponding factor $f_\ro{d}^{-1}$. The \lya emitters show
typical signs of radiative transfer effects, like large
velocity widths and asymmetric line profiles 
\citep[][Section 6.3]{2008ApJ...681..856R}, and should thus be optically
thick to \lya radiation. However, we have no good handle here on the actual
column density required to scatter emitted \lya at large radii effectively. It may well be lower than that required for damped
absorption, in which case it is very plausible that the region for
\lya emission extends beyond that for damped \lya absorption. In the
right panel of Figure \ref{fig:plot1}, the cumulative incidence rate
is shown for three values of the duty cycle $f_\ro{d} = (1,~0.5,~0.2$) with 
$v_\ro{c,0} = 50 \kmsec$ and $\alpha = 3$. Our simple assumption of a
duty cycle for the \lya emission reconciles the cumulative incidence
rate predicted by our model with the observed distribution for sizes
below $35$ kpc for $f_\ro{d} = 0.2$ ($f_\ro{d} = 0.4$ for $v_\ro{c,0} =70 \kmsec$). The
sudden drop in the observed distribution at $r \sim 35$kpc is likely
attributable to the following two effects. The first is the surface
brightness limit of the instrument. Light from sources with large
radii may be too diffuse to be detected. The second is the effect of
searching a small survey volume. Large systems are rarer, so there
is a limit to the size of sources that can be expected to be found
within the rather small survey volume.


\section{Modelling the luminosity function of the \lya emitters}

\subsection{The contribution of \lya cooling radiation}

\citet{2008ApJ...681..856R} considered a number of astrophysical
origins for the \lya emission that they observe. They conclude that
the most likely mechanism for producing \lya photons is star
formation. We here
consider in more detail the argument that cooling radiation is
unlikely to be the dominant source of \lya for the observed emitters.

\citet{2006ApJ...649...14D} derive the following formula for the \lya
luminosity $(L^\ro{c}_{\lya })$ of a collapsing protogalaxy due to cooling radiation,
 assuming that the gravitational binding energy
is radiated as \lya on a dynamical time scale,
\begin{equation} \label{eq:Llya} 
\begin{split} L^{\ro{c}}_{\lya } =& ~ 5.8 \times 10^{41}
\left(\frac{M_{\ro{tot}}}{10^{11}}\right)^{5/3}
\left(\frac{v_{\ro{amp}}}{v_{\ro{c}}}\right)
\left(\frac{1+z_{\ro{vir}}}{4}\right)^{5/2} \\ & \quad \times
\left(\frac{2-\alpha_{\ro{d}}}{2.5}\right)^{1.2} \ergs
\end{split}
\end{equation} where $M_{\ro{tot}}$ is the total (dark matter +
baryons) mass of the halo, $z_{\ro{vir}}$ is the redshift at which the
system virialises, and the bulk velocity of the infalling material,
$v_{\ro{bulk}}(r)$, is parameterised by $v_{\ro{amp}}$ and
$\alpha_{\ro{d}}$ as a power law, $v_{\ro{bulk}}(r) = v_{\ro{amp}}
(r/r_{\rm vir})^{\alpha_{\ro{d}}}$, where $r_{\rm vir}$ is the virial
radius.\footnote{See Equation (10) of \citet{2006ApJ...649...14D} for
a correction to this formula when $\alpha_{\ro{d}} < 0$ and $r$ is
small.} We will set $z_{\ro{vir}} = 3$, $v_{\ro{amp}} = v_{\ro{c}}$
and consider the lower limit of the range of $\alpha_{\ro{d}}$
discussed in \citet{2006ApJ...649...14D}, namely $\alpha_{\ro{d}} =
-0.5$, so that we have an upper limit on $L^{\ro{c}}_{\lya }$.

Equation \eqref{eq:Llya} gives a relation between the mass of a halo and
the luminosity due to \lya cooling, which we can combine with the \ps
formalism of Section \ref{dlamodel} to predict the number of DLAs per
unit comoving volume with \lya luminosity greater than some
$L^{\ro{c}}_{\lya }$,
\begin{equation} \label{eq:ngL} n(>L^{\ro{c}}_{\lya },z) =
\int_{M(L^{\ro{c}}_{\lya })}^\infty n_M(M',z) \df M'
\end{equation}

\begin{figure*} \centering
\begin{minipage}[c]{0.49\textwidth} \centering
\includegraphics[width=\textwidth]{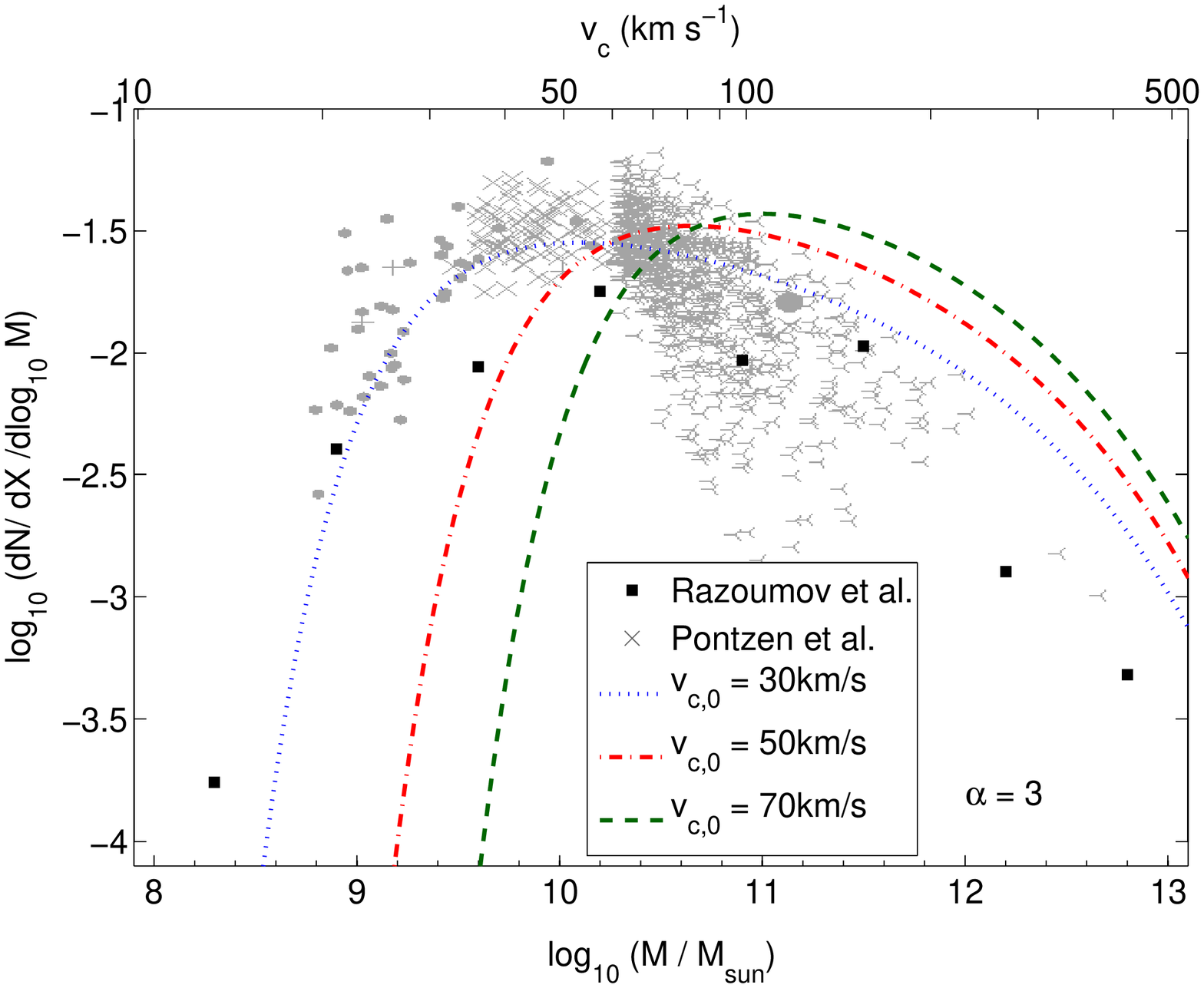}
\end{minipage}
\begin{minipage}[c]{0.49\textwidth} \centering
\includegraphics[width=\textwidth]{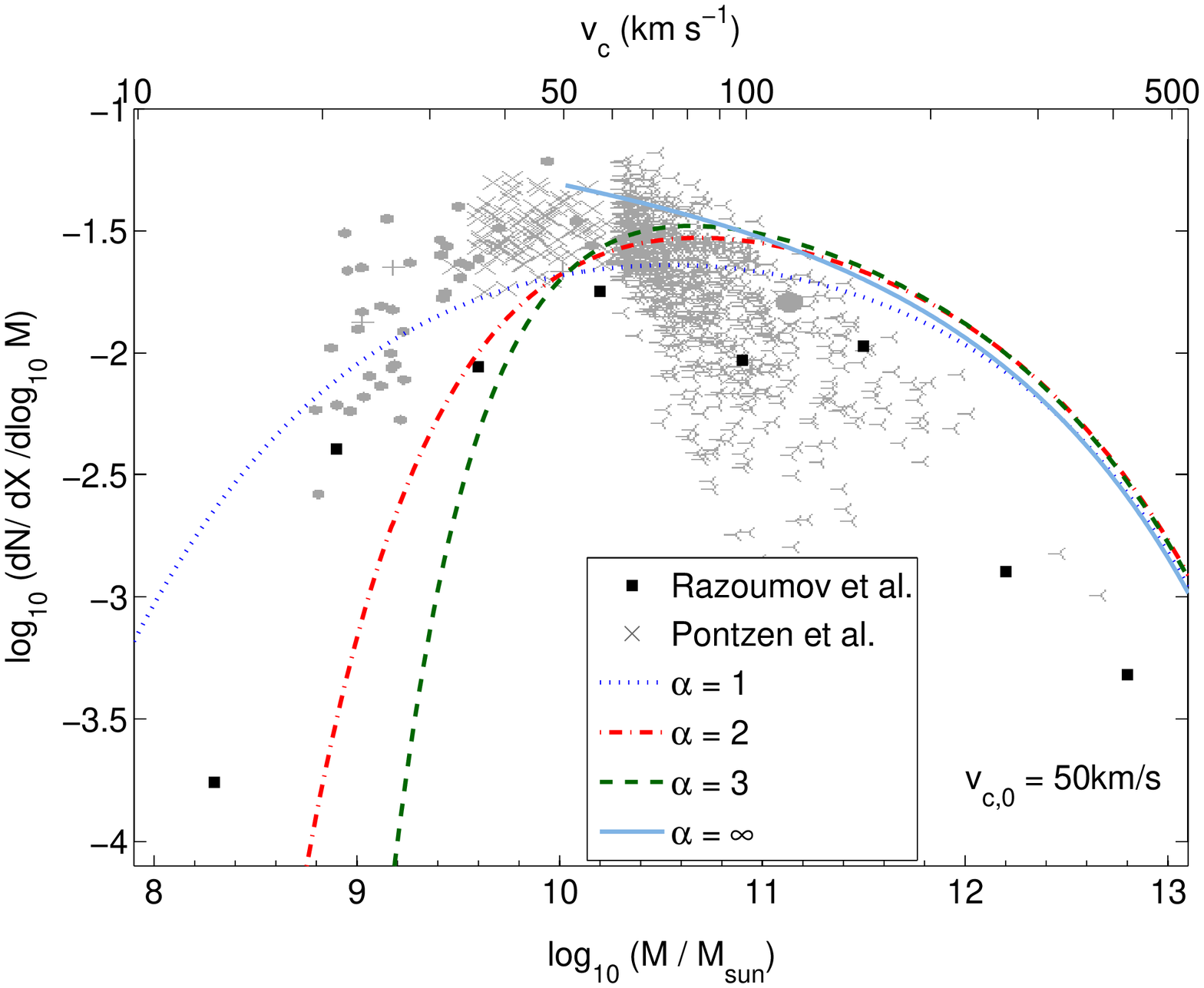}
\end{minipage}
\caption{The contribution of different mass ranges and virial velocity
ranges to the incidence rate of DLAs ($\dd^2 \cN / \dd X / \dd
\log_{10} M$). The black squares in both panels show the results from
the numerical model of \citet{2007arXiv0710.4137R} for DLAs found
within DM halos, excluding intergalactic DLAs. The grey
symbols in both panels show the results of the simulations of
\citet{2008arXiv0804.4474P}. \emph{Left:} The curves are for our model
with $\alpha = 3$ and a range of values of $v_\ro{c,0} = 30, 50, 70
\kmsec$. \emph{Right:} The curves are for our model with $v_\ro{c,0} = 50
\kmsec$ and a range of values of $\alpha = 1,2,3, \infty$.}
\label{fig:plot2}
\end{figure*}

There is, however, an inconsistency here. As already mentioned,
Equation \eqref{eq:Llya} assumes that the cooling radiation will be
emitted over the dynamical time $t_{\ro{dyn}}$ of the halo,
\begin{equation} t_{\ro{dyn}} \approx \frac{r_v}{v_{\ro{c}}} = \left(
\frac{1}{\frac{4}{3} \pi G \Delta_{\ro{v}} \bar{\rho}_m(z)}
\right)^{1/2} \approx 353 \ro{ Myrs}.
\end{equation} The proper time corresponding to the redshift interval
$[3.75, 2.67]$ of the \lya survey is $t_z = 789$ Myrs. Thus, all the
halos cannot radiate at the luminosity $L^{\ro{c}}_{\lya }$ given by Equation
\eqref{eq:Llya} for the entire redshift interval over which observations
were taken. There are two ways to make the model consistent. The
first is to impose a duty cycle, so that at any one time, only a
fraction $f_\ro{d} = t_{\ro{dyn}} / t_z \approx 0.45$ of the halos will be
radiating. 
The second is to reduce the luminosity of each
emitter\footnote{It could, of course, be a combination of these two
effects.}, so that the gravitational energy of the halo is radiated
over $789$ Myrs.

The left panel of Figure \ref{fig:cool} compares our prediction of $n(>L^{\ro{c}}_{\lya
},z)$ with the observed luminosity function
as given in Figure 9 of \citet{2008ApJ...681..856R}, shown in
black. The red dot-dashed curve shows the model assuming a duty cycle. The
green dashed curve is for the reduced luminosity. 

So far we have assumed \citep[following][]{2006ApJ...649...14D} that all
gravitational binding energy is converted into \lya radiation. In
reality, hot accretion flows could result in substantial amounts of
bremsstrahlung and He$^+$ line emission \citep{2005MNRAS.363....2K},
at the expense of $L^{\ro{c}}_{\lya }$. Furthermore, the observed
luminosities have not been corrected for slit losses which can be as
large as a factor of a few. We therefore conclude that 
cooling radiation is indeed unlikely to contribute significantly to
the majority of the faint \lya emitters. This further supports the suggestion by 
\citet{2008ApJ...681..856R} that the emitters are predominantly powered 
by centrally concentrated star formation surrounded by extended \lya halos. By stacking the
spectra of a subset of the emitters, \citet{2008ApJ...681..856R}
showed that the emission from these halos appears to extend to radii 
even larger than those for the individually detected emission plotted 
in Fig. 2 by a factor of at least two.


\subsection{ A simple model for the \lya luminosity function }

We will demonstrate now that a simple model where the
\lya luminosity due to stars $L^{\ro{s}}_{\lya }$ is proportional to
the total mass ($M \propto v_{\ro{c}}^3$) of the halos with virial velocity above a threshold
$v_{\ro{min}}$,
\begin{equation} \label{eq:Llyas} L^{\ro{s}}_{\lya } = \begin{cases}
L^{\ro{s}}_{0} \left( \frac{v_{\ro{c}}} {100 \kmsec} \right)^3 \ergs &
\text{if $v_{\ro{c}} > v_{\ro{min}}$}\\ 0 & \text{otherwise},
\end{cases}
\end{equation} fits the observed luminosity function remarkably
well. Note that detailed numerical simulations of much brighter \lya
break galaxies and \lya emitters appear to be consistent with this
simple scaling of the \lya emission with the properties of the DM host
halo \citep{2008arXiv0802.0228N}. In the right panel of Figure
\ref{fig:cool}, we compare the observed luminosity function of the
faint \lya emitters with a luminosity function modelled in this way
for a range of values of $f_\ro{d}$. The observed luminosity function is fit
well for the following parameter combinations $(f_\ro{d},L^{\ro{s}}_{0},
v_{\ro{min}} [\kmsec]) = (1, 4 \times 10^{41}, 75), (0.5,8 \times
10^{41}, 60), (0.2, 1.8 \times 10^{42}, 45)$. The upper and lower
boundary of the grey shaded region can be fit with values of 
$L^{\ro{s}}_{0}$ and $v_{\ro{min}}$ that differ from the quoted values 
by $\sim 25 \%$ and $\sim 5 \%$ respectively.

Our assumed scaling $L^{\ro{s}}_{\lya } \propto M$ is 
shallower than would be inferred from the star formation in 
most models of galaxy formation. The simulations of \citet{2008arXiv0804.4474P}
would {e.g.} predict $L^{\ro{s}}_{\lya } \propto M^{1.6}$. These
models are, however, generally tuned to produce a rather shallow
faint end of the luminosity function. As discussed by 
\citet{2008ApJ...681..856R}, their observations imply that the luminosity function of
\lya emitters steepens considerably at very faint luminosities \citep[cf.][]{2006MNRAS.365..712L}. 

Intriguingly, the values for the velocity cut-off and duty cycle in
our model are very similar to those required to fit the velocity width distribution
of the associated low-ionization metal absorption of DLAs and the
size distribution inferred from the \lya emitters. We caution,
however, that the significance of the apparent turn-over at $\sim 1.25
\times 10^{40} \ergs$ at the faint end of the observed luminosity
function is uncertain. At faint flux levels the luminosity function
will be affected strongly by the sensitivity limit of the
observations.

\section{The masses and virial velocities of DLA/LL host galaxies} If
the population of faint \lya emitters detected by
\citet{2008ApJ...681..856R} can be identified with DLA/LL host
galaxies, then this constitutes the first measurement of the space
density and average size of DLA/LL host galaxies. The last section
supported the suggestion by \citeauthor{2008ApJ...681..856R} that the
\lya emission is powered by star formation. In this case, with
standard assumptions for the conversion of \lya emission to star
formation rate, the \lya luminosities correspond to star formation
rates of 0.07 - 1.5 \Msol yr$^{-1}$, similar to that inferred by
\citet{2003ApJ...593..215W} from the CII* $\lambda$1335.7
absorption in DLAs. No continuum is detected, so
there is no information about stellar or total masses of the
object. Our modelling is thus currently the only handle we have on the
masses (and virial velocities) of what should be a statistically
representative sample of DLA host galaxies.

In Figure \ref{fig:plot2}, we show the contribution of DM halos of
different masses and virial velocities to the incidence rate of DLAs
in our model, for the range of parameters used to model the
suppression of the cross-section for damped \lya absorption in
low-mass DM halos. We also show the results from 
two recently published numerical simulations of DLAs
\citep{2007arXiv0710.4137R,2008arXiv0804.4474P}. The differential line
density for DLAs is calculated similarly to $N$, except that we
consider intervals of $\dd X$ and $\dd \log_{10}M$,
\begin{equation} \frac{\dd^2 \cN} {\dd X ~ \dd \log_{10} M} =
\frac{c}{H_0} \ln 10 ~ M ~ n_M(M,X) ~ \sigma(M,X).
\end{equation} Note that larger values of $\alpha$ result in a sharper
turnover at low masses, at the expense of increasing the abundance of
DLAs with high masses (the area under the plot is normalised).

The majority of the DLAs in our model have virial velocities in the
range 50 to 200 $\kmsec$, corresponding to total masses of $10^{10}$
to $10^{12} \Msol $. As discussed in the previous sections, the
turn-over at small virial velocities is most constrained by the
velocity width distribution of the associated low-ionization metal
absorption and is most likely attributable to feedback effects due to
star formation. The decline at large virial velocities and masses is
due to the decline of the space density of DM halos.

The incidence rate in the numerical simulations of
\citet{2007arXiv0710.4137R} shows a similar peak, albeit shifted to
somewhat smaller masses/virial velocities than our model requires to
fit the kinematical data of the DLAs. This is perhaps not surprising
--- \citet{2007arXiv0710.4137R} find that the velocity widths in their
simulations fall somewhat short of those observed. Their simulation
also takes into account DLAs that are not contained within any halo
i.e. intergalactic DLAs.

The numerical simulations of \citet{2008arXiv0804.4474P} show a
sharper peak centered on virial velocities of 30-80 $\kmsec$. 
Such a sharp peak at rather low virial velocities appears, however, at
odds with the observed velocity widths of the associated
low-ionization metal absorption. In the simulations of \citeauthor{2008arXiv0804.4474P}, the decline of the contribution to 
the incidence rate with increasing mass is much faster than the
decline of the space density of massive halos. This fast decline is 
due to a flattening of the absorption cross section with increasing 
mass in massive halos (\citeauthor{2008arXiv0804.4474P}, Figure 4). 


\section{Summary and Conclusions} We have considered an updated
version of the Haehnelt et al. model for the kinematics of DLAs, in
light of the discovery of a new population of extended low surface
brightness \lya emitters with a total inferred incidence rate similar
to that of DLAs. The main differences with the modelling of
\citet{2000ApJ...534..594H} are the use of the \st modification to the
\ps formalism, an update of cosmological parameters, and the use of an
exponential suppression of the cross section for damped absorption for
low virial velocities instead of an sharp cut-off.

Our main results are the following. 

\begin{itemize}

\item{The observed velocity width distribution of the associated metal
absorption can be fit with a model where the cross section for damped
absorption scales with the virial velocity of the halo as $\sigma
\propto v_\ro{c}^{2.5}$, the absorption cross section is suppressed in
halos with $ v_\ro{c}\le 70\kmsec$, and the conditional velocity width
is given by that in the simulations of Haehnelt et al. (or the very
similar distribution of the simulations of
\citealt{2008arXiv0804.4474P}).}

\item{The same model can fit the cumulative incidence rate for
absorption inferred from the size distribution of the \lya emitters if
the \lya emission has a duty cycle of $f_\ro{d} = 0.2-0.4$, and the emission
extends over an area which is larger than the cross section for damped
\lya absorption by a factor $f_\ro{d}^{-1} = 2.5-5$.} 

\item{The maximum expected \lya cooling luminosity due to collapsing gas in DM
halos falls short of the observed \lya luminosities by a factor of
three to five, even for optimistic assumptions regarding the expected
emission due to \lya cooling. Furthermore, the expected dependence of the \lya
cooling luminosity on the virial velocity of DM halos thereby maps into a
luminosity function with a slope shallower than observed. \lya cooling
radiation should thus not contribute significantly to the \lya
emission for the majority of the objects, especially at the faint end
of the \lya luminosity function.}

\item{The \lya luminosity function is well fit by a simple model where
the \lya luminosity scales linearly with the mass of the DM halo and the
emission is suppressed for low mass DM halos. 
The \lya luminosity function can be fit for a wide range of duty
cycles including the duty cycle required to simultaneously explain the kinematic
properties of DLAs and the cumulative incidence rate inferred from the
observed size distribution of the \lya emitters.}

\item{Our model predicts that the bulk of the contribution to the
incidence rate of DLAs comes from absorption systems hosted in DM
halos with virial velocities in the range from 50-200 $\kmsec$ and
masses in the range $10^{10}$ to $10^{12} \Msol$. The cut-off 
for damped absorption occurs at somewhat higher virial velocities 
than suggested by numerical simulations, which attempt
to simulate the gas distribution and kinematics of DLAs
self-consistently. These simulations, however, fall short of
reproducing the observed velocity widths of the associate
low-ionization metal absorption of DLAs. If the suppression of the
cross section for damped \lya absorption in halos with virial
velocities up to $70 \kmsec$ is indeed real, then feedback due to star
formation at high redshift has to be more efficient in removing gas
--- even from rather deep potential wells --- than is assumed in most
models of galaxies formation and numerical simulations. Alternatively 
the simulation (and our model) may underestimate the effect of stellar feedback 
on the velocity width of absorbers hosted by DM halos with small
virial velocities. This may not be implausible given the fact 
that the simulations still fail to reproduce realistic galactic
winds.}
\end{itemize}

Our modelling here adds to the mounting evidence that DLAs and LLs
predominantly probe the outer regions of dwarfish\footnote{``In
English the only correct plural of dwarf is dwarfs, and the adjective
is dwarfish \ldots dwarves and dwarvish are used \ldots only when
speaking of the ancient people to whom Thorin Oakenshield and his
companions belonged.'' \citet{Tolkien}.} galaxies. This picture had
been suggested for a while by the kinematic and metallicity data. With
the discovery of a faint population of \lya emitters, most plausibly
identified with the population of DLA/LL host galaxies, we finally
have a handle on their space density, sizes and (in conjunction with
models like the one presented here) masses and virial velocities. In
the picture that emerges, DLAs are hosted by the galaxies that, in the
context of the now well established $\Lambda$CDM paradigm for
structure formation, are expected to become the building blocks of
typical galaxies like our Milky Way.


\section*{Acknowledgments} LAB is supported by an Overseas Research
Scholarship and the Cambridge Commonwealth Trust. We thank Andrew Pontzen 
providing $p(v_\ro{w}| v_{\ro{c}}) \df v_\ro{w}$ from his simulations and giving useful comments 
on a draft of this paper. Max Pettini, George Becker and Michael Rauch also deserve thanks for 
comments and useful discussion.

\bsp 
\label{lastpage}

\end{document}